\documentclass[12pt, dvipdfmx]{revtex4}

\usepackage{amsmath}
\usepackage{amssymb}
\usepackage[dvips]{graphicx,color}

\newcommand{\beq}{\begin{eqnarray}}
\newcommand{\eeq}{\end{eqnarray}}
\def\tr{\mathop{\mathrm{tr}}\nolimits}

\begin{document} 

\title{Lattice study on QCD-like theory with exact center symmetry}

\author{Takumi Iritani}
\email{iritani@yukawa.kyoto-u.ac.jp}
\affiliation{Yukawa Institute for Theoretical Physics, Kyoto 606-8502, Japan}

\author{Etsuko Itou}
\email{eitou@post.kek.jp}
\affiliation{High Energy Accelerator Research Organization (KEK), Tsukuba 305-0801, Japan}

\author{Tatsuhiro Misumi}
\email{misumi@phys.akita-u.ac.jp}
\affiliation{Department of Mathematical Science, Akita University, 
1-1 Tegata Gakuen-machi, Akita 010-8502, Japan\\
Research and Education Center for Natural Sciences, 
Keio University, 4-1-1 Hiyoshi, Yokohama, Kanagawa 223-8521, Japan}

\begin{abstract}
We investigate QCD-like theory with exact center symmetry, with emphasis on
the finite-temperature phase transition concerning center and chiral symmetries.
On the lattice, we formulate center symmetric $SU(3)$ gauge theory with 
three fundamental Wilson quarks by twisting quark boundary conditions 
in a compact direction ($Z_3$-QCD model).
We calculate the expectation value of Polyakov loop and the chiral condensate 
as a function of temperature on $16^3 \times4$ and $20^3 \times 4$ lattices 
along the line of constant physics realizing $m_{PS}/m_{V}=0.70$.
We find out the first-order center phase transition, 
where the hysteresis of the magnitude of Polyakov loop exists 
depending on thermalization processes.
We show that chiral condensate decreases around the critical temperature 
in a similar way to that of the standard three-flavor QCD,
as it has the hysteresis in the same range as that of Polyakov loop. 
We also show that the flavor symmetry breaking due to the twisted boundary condition 
gets qualitatively manifest in the high-temperature phase.
These results are consistent with the predictions based on the chiral effective model in the literature.
Our approach could provide novel insights to the nonperturbative connection between 
the center and chiral properties.
\end{abstract}

\maketitle

\newpage

\section{Introduction}
\label{sec:intro}

Strong dynamics based on Quantum ChromoDynamics (QCD) induces two 
main nonperturbative phenomena, ``quark confinement" and 
``spontaneous chiral symmetry breaking", which are the main themes of
research on strong-coupling dynamics.
While both of the two phenomena are caused by infrared physics based on the asymptotic freedom,
it is not fully understood how they are related with each other.
Some of lattice QCD simulations show that these two phenomena are almost simultaneously 
turned off at a certain temperature~\cite{Kogut:1982rt,Fodor:2009ax,Bazavov:2013txa},
where the transition to the quark-gluon plasma (QGP) phase occurs.
Regarding the connection between the two phenomena,
we here have several questions:
Do the transitions take place at the same temperature?
Does one of the transitions induce the other?
Investigating these questions could lead to understanding of the nonperturbative 
relation between confining and chiral symmetry breaking.
There are several approaches for study on the linkage between these phenomena,
including the analysis on quark confinement in terms of chiral symmetry breaking~\cite{Gongyo:2012vx, Iritani:2013pga, Doi:2014zea,Gattringer:2006ci,Bruckmann:2006kx,Synatschke:2007bz,Synatschke:2008yt,Bilgici:2008qy}
or vise versa~\cite{Miyamura:1995xn,Woloshyn:1994rv} from lattice QCD,
and the effective models motivated by both chiral and center symmetries~\cite{ Hatta:2003ga,Fukushima:2003fw,Sasaki:2006ww,Ratti:2006wg,Fukushima:2008wg}.

In the presence of exact $Z_{3}$ center symmetry,
one can study the confining/deconfining phase transition 
by use of the expectation value of the Polyakov loop.
Since it is the order parameter
of $Z_{3}$ center symmetry, 
it works as an indicator of confining/deconfining through the quark free energy.
Quenched QCD (without dynamical quarks) \cite{Kogut:1982rt} and adjoint QCD 
(with adjoint quarks instead of fundamental ones) \cite{Karsch:1998qj} are examples of this case.
However, their results on coincidence
of the transition temperatures are inconsistent:   
For the quenched QCD it is reported that the transition temperatures coincides
while they do not for the adjoint QCD.
In the first place, these two theories are too far from the realistic QCD in a sense that
one contains no dynamical quarks while the other contains an excessive amount of 
dynamical degrees of freedom of quarks.

For the physical $N_{f}=2+1$ QCD, where $u$- and $d$-quark masses are degenerated while $s$-quark has a heavier mass, things get more involved:
the dynamical fundamental quarks explicitly break 
center symmetry, so what we can do the best
is study the crossover transition based on the expectation values of the Polyakov loop.
In the literature \cite{Bazavov:2013txa,Fodor:2009ax}, one finds inconsistent results 
on coincidence of the crossover temperatures of center and chiral
transitions, depending on the scheme and setup of the lattice simulations:
Some works argue that the two 
critical temperatures are very close \cite{Bazavov:2013txa} while the others show
that they are not \cite{Fodor:2009ax}.

One ideal situation for studying this topic in details is 
that the {\it exact center symmetry} and the {\it dynamical fundamental quarks} are reconciled.
As well-known, the existence of dynamical fundamental quarks breaks center symmetry explicitly,
thus the above accommodation seems infeasible.
However, appropriate boundary conditions for quarks enable us to realize such a situation.
By imposing three different twisted boundary conditions on the three fundamental
quarks (shifted by $2\pi/3$) in the compact imaginary-time direction, 
we realize center symmetric $SU(3)$ gauge theory with
three dynamical quarks in fundamental representation on $R^3 \times S^1$ ($Z_{3}$-QCD model)
\cite{Kouno:2012zz, Sakai:2012ika, Kouno:2013zr,Kouno:2013mma,Kouno:2015sja}.
By investigating this model, we could make progress in elucidating the connection 
between confining/deconfining and chiral transitions.
The analytical study on this theory based on the Polyakov-loop extended chiral model 
\cite{Fukushima:2003fw,Ratti:2006wg,Fukushima:2008wg}
was initiated in Ref.~\cite{Kouno:2013mma}, 
which shows that the center symmetry is spontaneously broken
at certain temperature, associated by the manifestation of flavor symmetry breaking.
Although the chiral model helps understand rough picture of the nonperturbative properties,
we cannot fully eliminate the model artifacts.

In this paper, we numerically investigate finite-temperature $Z_{3}$-QCD model on the lattice, 
with emphasis on center phase transition and its influence on the chiral properties.
We formulate center symmetric $SU(3)$ lattice gauge theory with 
three fundamental Wilson quarks by twisting the boundary conditions 
in the imaginary time direction.
We calculate the expectation value of Polyakov loop 
and the chiral condensate on $16^3 \times 4$ and $20^3 \times 4$ lattices 
with $m_{PS}/m_{V}=0.70$ fixed.
We find out the first-order center phase transition at certain temperature,
where the hysteresis exists depending on the initial condition for the configuration generation.
We show that the chiral condensate rapidly decreases around the center critical temperature 
and has the hysteresis too.
We also verify manifestation of the flavor symmetry breaking in the flavor-diagonal meson sector 
in the high-temperature phase, which was predicted in the effective chiral model study.

The structure of the paper follows:
In Sec.~\ref{sec:model}, we briefly review the construction of $Z_{3}$-QCD model in the continuum theory and  
introduce the expected properties based on the chiral model study. 
In Sec.~\ref{sec:setup}, we formulate $Z_3$-QCD model on the lattice and note details of the simulation setup
including the parameter fixing for finite-temperature simulations.
In Sec.~\ref{sec:results}, we show our results for the phase transitions associated with the center and chiral symmetries.
Sec.~\ref{sec:summary} is devoted to summary and discussion.

\section{$Z_3$-QCD model}
\label{sec:model}

\subsection{Formulation as a continuum theory}
We give a brief review of $Z_{3}$ center-symmetric $SU(3)$ gauge theory with three fundamental quarks on $R^{3} \times S^1$ proposed in Refs.~\cite{Kouno:2012zz}. 
Here, the compact dimension can be seen as imaginary time direction in the present work.  
We first consider the partition function $Z$ in Euclidian spacetime with one compact dimension as
\begin{align}
&Z\,=\, \int D\Psi D\bar{\Psi} DA\,\exp[-S],
\\
&S\,=\, \int d^{3}x \int_{0}^{\beta} d\tau 
\left[
\sum_{f}\bar{\Psi}_{f}(\gamma_{\mu}D_{\mu}+m)\Psi_{f} \,+\, {1\over{2g^{2}}}\tr F^{2}_{\mu\nu}
\right]\,,
\label{Action}
\end{align}
where $\Psi_{f}$ is a quark field for degenerate three-flavor quarks with flavor index $f=1,2,3$.
$D_\mu \equiv \partial_\mu-iA_\mu$ is a covariant derivative with the $SU(3)$ gauge field $A_{\mu}$
and the field strength is given by $F_{\mu\nu} =\partial_{\mu}A_{\nu}-\partial_{\nu}A_{\mu}-i[A_{\mu},A_{\nu}]$.

We here regard a compact dimension ($\tau$) as imaginary time direction, 
and impose an anti-periodic boundary condition on a quark field in the direction as
\begin{equation}
\Psi_f ( \vec{x}, \tau =\beta)=-\Psi_f (\vec{x},\tau =0)\,.
\label{ABC}
\end{equation}
By adopting this boundary condition, we now work on the finite-temperature system with $\beta=1/T$.

It is known that the above action itself is invariant under the $Z_{3}$ center transformation, 
which is generated by the center elements of global color $SU(3)$ transformation.
However, this transformation eventually gives rise to $2\pi /3$ shift of phase of the quark boundary condition as
\begin{equation}
\Psi_f (\vec{x},\tau =\beta)=-e^{i2\pi k/3}\,\Psi_f (\vec{x}, \tau =0)\,,
\label{Z3tr}
\end{equation}
with $k=0,1,2$.
Thereby, the $Z_{3}$ center symmetry is explicitly broken via the fundamental-quark boundary conditions 
(\ref{ABC})(\ref{Z3tr}) in the usual three-flavor $SU(3)$ gauge theory, and of course 
in the realistic $N_{f}=2+1$ QCD.

We now consider a case that the three flavors have distinct boundary conditions as following,
\begin{align}
\Psi_f (\vec{x}, \tau =\beta)=- e^{i 2 \pi (f-1)/3}\Psi_f(\vec{x}, \tau =0), 
\label{Z3BC}
\end{align}
for $f=1,2,3$.
It is nothing but $SU(3)$ gauge theory in the presence of
three fundamental quarks with flavor-dependent boundary conditions. 

For this case, the $Z_{3}$ transformation in Eq.~(\ref{Z3tr}) shifts the quark boundary conditions as
\begin{align}
 \Psi_f(\vec{x}, \tau =\beta)=-e^{i2\pi (k+f-1)/3}\Psi_f(\vec{x},\tau =0).
\label{Z3BC2}
\end{align}
The twisted angles for the three flavors are changed into $k+f-1$,
but this can be straightforwardly relabeled as $f-1$ and returns back to the first place in Eq.~(\ref{Z3BC}).
In other words, we can rename the flavor $f+k$ as $f \pmod{3}$.
This means that the three-flavor $SU(3)$ gauge theory with the special twisted boundary condition
Eq.~(\ref{Z3BC}) is invariant under the $Z_{3}$ center transformation.
We call this exactly-center-symmetric model as {\it $Z_{3}$-QCD model}
\cite{Kouno:2012zz, Sakai:2012ika, Kouno:2013zr,Kouno:2013mma,Kouno:2015sja}.
We note that the flavor-dependent twisted boundary condition is translated 
into the insertion of the flavor-dependent
imaginary chemical potential by use of gauge transformation as shown in Appendix.~\ref{sec:Imch}.
Hereafter, we use $f=u,d,s$ instead of $f=1,2,3$ as indices for flavor.

\subsection{Chiral structure and flavor symmetry in $Z_3$-QCD model}\label{sec:chiral-Z3}
We here comment on flavor symmetry possessed by $Z_{3}$-QCD model in the massless limit.
In $Z_{3}$-QCD, due to the boundary conditions, the flavor-chiral $SU(3)_{L}\times SU(3)_{R}$ 
symmetry is explicitly broken to its Cartan subgroup $U(1)_{L}^{2} \times U(1)_{R}^{2}$ 
generated by $\lambda_{3}, \lambda_{8}$ of the Gell-mann matrix elements.  
However, this influence from the boundary condition disappears in the infinite compact-circumference limit,
or in the zero-temperature limit.
Thus, in this limit, $Z_{3}$-QCD is reduced to the standard three-flavor QCD at zero temperature.

For nonzero temperature, the effects of the twisted boundary condition exists in principle, namely 
the flavor-chiral symmetry is broken to $U(1)_{L}^{2} \times U(1)_{R}^{2}$ at the action level. 
In this work, we regard this Cartan subgroup 
as the specific flavor-chiral symmetry of the $Z_3$-QCD model.
Thus, the pattern of the chiral symmetry breaking is basically given by 
$U(1)^2_L \times U(1)^2_R \rightarrow U(1)^2_{\tilde{V}}$ in this model. 
While the chiral transition in the standard three-flavor QCD model is believed to be first-order in the chiral limit
based on the universality class~\cite{Pisarski:1983ms}, the same discussion is probably not valid for the $Z_3$-QCD model.

Here, we make a supplemental comment on the flavor-chiral symmetry 
$U(1)_{L}^{2} \times U(1)_{R}^{2}$ in the $Z_{3}$-QCD model discussed above.  
It is notable that, indeed, we do not know at how high temperature the physical quantities 
start to be affected by the twisted boundary condition. 
It may be as soon as the temperature is turned on or may be 
at a certain nonzero temperature.
Actually, as we will see in the next subsection, the study based on the chiral effective model indicates that the chiral condensate is insensitive to the boundary condition below the center phase transition temperature,
where the flavor symmetry breaking due to the boundary condition 
is not manifest in the effective thermodynamic potential.

The significance of this model is that we can study the phase transition with respect to the $Z_{3}$ 
center symmetry by calculating expectation values of Polyakov loop, 
even in the presence of the dynamical fundamental quarks.
We here have physical and theoretical questions:
If the phase transition occurs, what is the order of the phase transition?
How is the chiral condensate affected by the center phase transition?
We study these topics in the next section numerically,
although our simulations are not carried out in the chiral limit.
The study on such an ideal model with the exact center symmetry
may help elucidate the relation of center 
and chiral properties from novel viewpoints.

\subsection{Expected properties from chiral model and QCD with a finite chemical potential}
\label{subsec:chiM}

In the study of $Z_{3}$-QCD based on the Polyakov-loop-extended NJL (PNJL) model~\cite{Fukushima:2003fw,Ratti:2006wg,Fukushima:2008wg},
the following three properties are predicted~\cite{Kouno:2013mma} (we call this model just as the effective chiral model in the present paper.):

(i) $Z_{3}$ center symmetry is spontaneously broken in the high-temperature phase,
where the order of the phase transition is first.

(ii) Although the $SU(3)$ flavor symmetry the standard three-flavor QCD has is spoiled by the twisted boundary condition at the action level, 
the effective thermodynamical potential of the PNJL model in the low-temperature phase 
is not affected by the boundary condition. It is expected that the $Z_3$-QCD model becomes $SU(3)$ flavor-symmetric in the phase.
On the other hand, the flavor symmetry breaking becomes manifest in the high-temperature phase.

(iii) In a chiral limit, even above the critical temperature of the center symmetric phase transition,
the chiral condensate has a nonzero value and the chiral symmetry is still broken. 
It might be an artifact coming from the model cutoff in the chiral effective model.
Instead, the value of chiral condensate has a specific jump at the center critical temperature.

The present model is also related to the works on QCD with finite imaginary chemical potential.
As discussed in Refs.~\cite{Roberge:1986mm, D'Elia:2002gd},
the partition function has $2\pi/3$ periodicity in the imaginary chemical potential.
Thus we speculate that the critical temperature for the chiral phase transition in the present model, 
which also has $2\pi /3$ periodicity of the temporal direction,
could be the same as that of the standard three-flavor QCD.

\section{Simulation setup}\label{sec:setup}
We utilize the Iwasaki gauge action with naive Wilson fermions in our lattice numerical simulation.
The definition of the action is given by
\beq
S&=&S_g + S_f, \nonumber\\
S_g &=& \beta \sum_x \left( c_0 \sum^4_{\mu < \nu; \mu,\nu=1} W_{\mu \nu}^{1 \times 1} (x) +c_1 \sum^4_{\mu \ne \nu;\mu,\nu=1} W^{1 \times 2}_{\mu \nu} (x)   \right),\\
S_f&=&\sum_{f=u,d,s} \sum_{x,y} \bar{\psi}^f_x M_{x,y} \psi^f_y,\label{eq:fermi-action}
\eeq
where $\beta=6/g^2$, in which $g$ is a lattice bare gauge coupling constant, $c_1=-0.331$, $c_0=1-8c_1$ and
$W^{1\times 1}$ and $W^{1 \times 2}$ denote the plaquette and rectangular, respectively. 
In Eq.~(\ref{eq:fermi-action}),
\beq
M_{x,y}&=&\delta_{x,y} -\kappa \sum_{\mu=1}^4 \left\{ (1-\gamma_\mu) U_{x,\mu}\delta_{x+\hat{\mu},y} + (1+\gamma_\mu) U^\dag_{y,\mu} \delta_{x,y+\hat{\mu}}  \right\}.\label{eq:def-Dirac}
\eeq
Here $\kappa$ is the hopping parameter. 
We note that the value of $\kappa$ in $Z_3$-QCD model is universal for all flavors.

To realize the twisted boundary condition in Eq.~(\ref{Z3BC}) on the lattice, we introduce the following boundary conditions for the link variable only in the fermion action Eq.~(\ref{eq:fermi-action}):
\beq
U_4 (\vec{x}, \tau=N_\tau) =& -  U_4 (\vec{x}, \tau=0) & \mbox{ for $u$-flavor }, \nonumber \\
U_4 (\vec{x}, \tau=N_\tau) =& - e^{2 \pi i/3} U_4 (\vec{x}, \tau=0) &\mbox{ for $d$-flavor}, \nonumber \\
U_4 (\vec{x}, \tau=N_\tau) =& -  e^{4 \pi i / 3}U_4 (\vec{x}, \tau=0) & \mbox{ for $s$-flavor}.
\eeq
These conditions from top to bottom are the same with those for the standard finite-temperature QCD with imaginary chemical potential; $\mu_I=0, 2\pi /3$ and $4 \pi/3$, respectively.
We have to use the Rational Hybrid Monte Carlo (RHMC) algorithm to calculate the fermion determinant for each flavor, since three fermions have a different boundary condition with each other.

Firstly, we perform the zero-temperature simulation using $16^4$ lattices to obtain the line of constant-physics.
We carry out the simulations for several values of hopping parameter $\kappa$ with each $\beta$ value and measure the flavor-singlet pseudo-scalar mass ($m_{PS}$) and vector meson mass ($m_V$) for each flavor.
The number of trajectories we generate is $2,000$ -- $3,000$, and we measure the correlator of these hadronic states every $10$ Monte Carlo trajectories. The estimated autocorrelation length is around $100$ trajectories in this simulation.
  We summarize results on mass measurement in Table~\ref{tab:mass_summary_1} and \ref{tab:mass_summary_2}
in Appendix~\ref{sec:mass_summary}.

\begin{table}[h]
\begin{center}
\begin{tabular}{|c|c|c|c|c|c|c|c|c|c|c|c|c|c|c|}
\hline
$\beta$    &$1.30$      & $1.40$       & $1.50$      & $1.55$      & $1.60$  & $1.70$     & $1.80$ & $1.90$ & $1.95$ & $2.00$ & $2.10$ & $2.20$ \\  
\hline
$\kappa$ & $0.2019$  &  $0.1975$ & $0.1921$  &  $0.1892$ & $0.1861$&  $0.1793$ & $0.1725$ & $0.1663$ & $0.1636$ & $0.1611$ & $0.1571$ & $0.1539$ \\
\hline
\end{tabular}
\caption{ Simulation parameters: $\beta$ and $\kappa$}  \label{table:sim-parameter}
\end{center}
\end{table}

We fix the ratio between $m_{PS}$ and $m_V$ constant, namely $m_{PS}/m_{V}=0.70$, and tune the value of hopping parameter for each $\beta$ shown in Table~\ref{table:sim-parameter}.
Using these parameter sets in Table~\ref{table:sim-parameter}, 
we perform the finite temperature simulation on $16^3 \times 4$ and $20^3\times 4 $.
The number of trajectories for the finite temperature lattice setup is $500$ -- $5,000$.
We measure the Polyakov loop in temporal direction for every Monte Carlo trajectory and the chiral condensate every $10$ trajectories.

We note that the masses of flavor-singlet mesons are the same with that of the standard three-flavor QCD in zero-temperature.
It shows that, at least in the flavor-singlet sector, the breaking of 
$SU(3)_{L}\times SU(3)_{R}$ flavor-chiral symmetry due to
the $Z_{3}$ twisted boundary condition (\ref{Z3BC}) is not observed for zero-temperature. 
It is consistent with the general argument in the previous section and 
the result of PNJL model in Sec.~\ref{subsec:chiM}.

\section{Simulation results}
\label{sec:results}

\subsection{Polyakov loops and center symmetry}
\begin{figure}
\centering
\includegraphics[width=0.47\textwidth,clip]{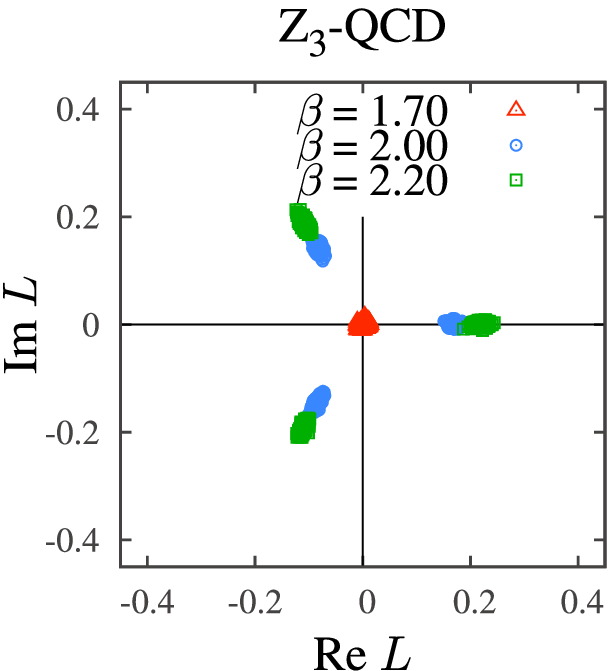} 
\includegraphics[width=0.47\textwidth,clip]{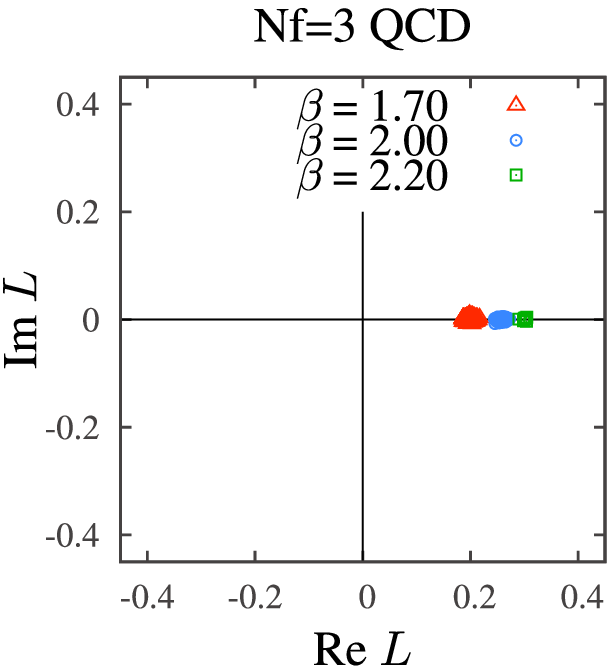}
\caption{Polyakov loop distribution plot in $Z_{3}$-QCD (left) and the standard three-flavor QCD (right).
Based on $16^3 \times 4$ lattice for $\beta=1.70,2.00,2.20$ with the same values of $\kappa$ in both panels.
}
\label{fig: Pdis1}
\end{figure}

We first show the existence of $Z_{3}$ center symmetry in the present model 
based on the distribution plot of Polyakov loop ($L$),
\beq
L= \frac{1}{V} \sum_{\vec{x}} \frac{1}{3} \tr \left[ \prod_{i=1}^{N_\tau} U_\tau (\vec{x},i) \right].
\eeq
Here $V$ denotes the spacial volume in a lattice unit.
As shown in the left panel of Fig.~\ref{fig: Pdis1}, the Polyakov loops are distributed around the origin
in the low $\beta$ regime while three vacua exist in the high $\beta$ regime for $Z_{3}$-QCD.
On the other hand, those in the standard three-flavor $SU(3)$ gauge theory 
in the right panel of Fig.~\ref{fig: Pdis1}
indicate explicit breaking of $Z_{3}$ center symmetry.
These results obviously show that the $Z_{3}$-QCD model possesses exact $Z_{3}$ center symmetry at the action level while it seems to undergo spontaneous breaking of the symmetry in the high-temperature phase.

\begin{figure}
\centering
\includegraphics[width=0.47\textwidth,clip]{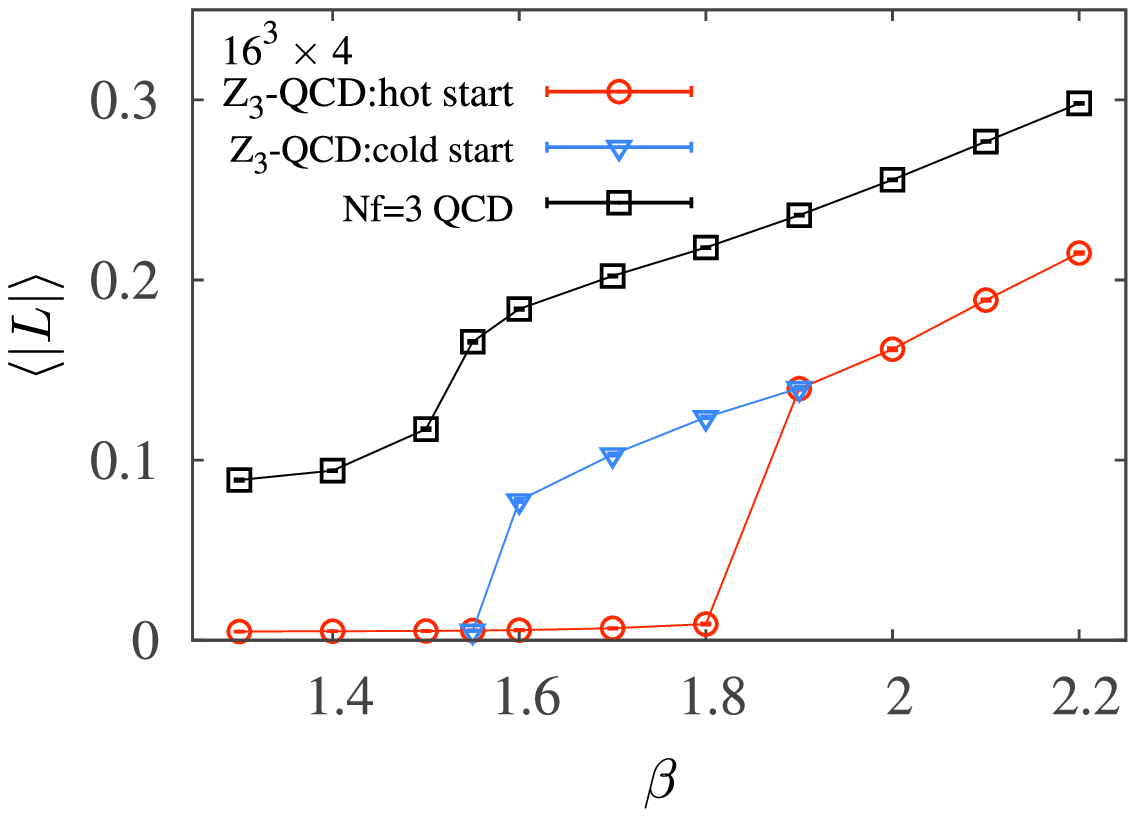} 
\includegraphics[width=0.47\textwidth,clip]{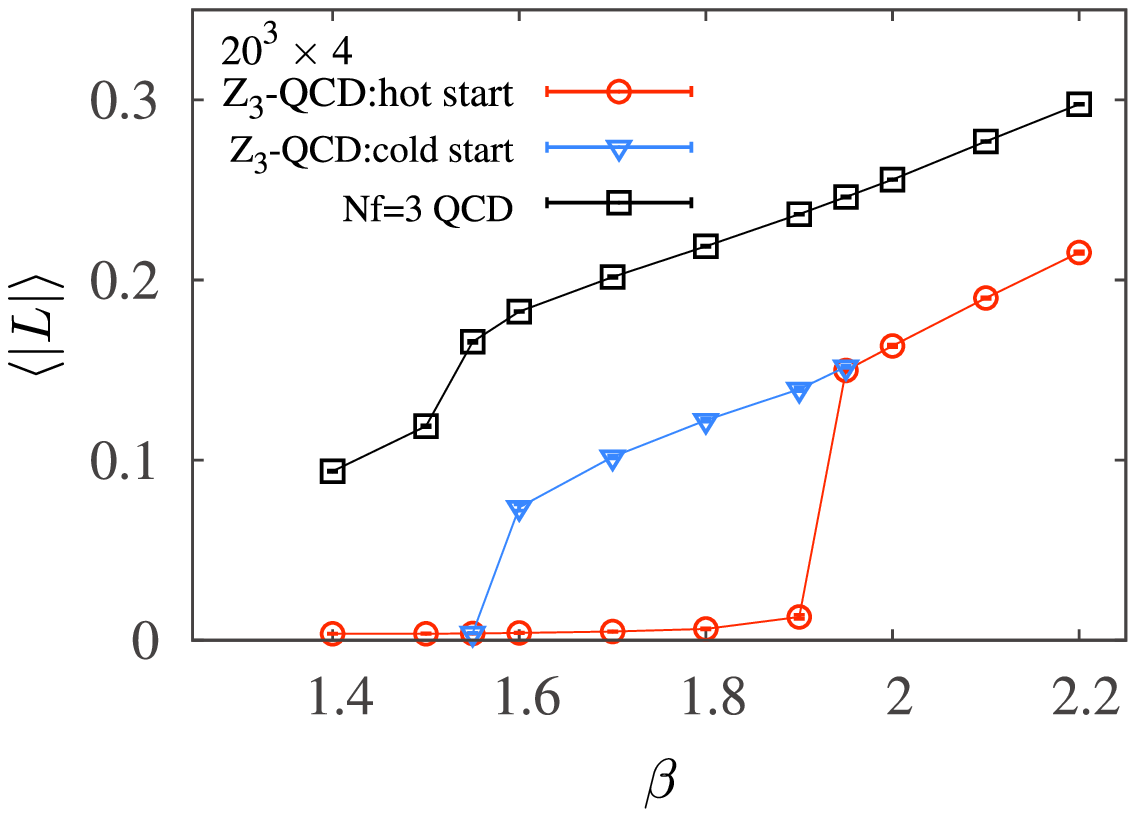} 
\caption{ 
$\beta$ dependence of the magnitude of Polyakov loop ($\langle |L| \rangle$)
for the $Z_{3}$-QCD and standard three-flavor QCD on $16^3 \times 4$ (left) and $20^3\times 4$ (right) lattices.
For the $Z_3$-QCD model, the data of $\langle |L| \rangle$ started with the cold start (triangle (blue) symbols) have a clear jump from zero to non-zero values around the region $1.55 \le \beta \le 1.60$ in both panels,
while the jump occurs in $1.80 \le \beta \le 1.90$ (left) and $1.90 \le \beta \le 1.95$ (right) for the data generated by the hot start (circle (red) symbols).
In the regions between these two jumps, the hysteresis exists in $Z_3$-QCD model.
On the other hand, the data of the standard three-flavor ($N_f=3$) QCD (square (black) symbols) do not show such a jump from zero to nonzero nor hysteresis.
}
\label{fig:Pb}
\end{figure}

Next, we investigate temperature dependence of the Polyakov loop by varying $\beta$
along with the line of constant-physics, namely $m_{\rm PS}/m_{V}=0.70$, shown in Table~\ref{table:sim-parameter}.
We generate configurations with two types of initial condition; cold start and hot start.
In both panels of Fig.~\ref{fig:Pb}, the triangle (blue) symbol denotes the data started with “cold start”. The corresponding initial configuration lives in the ordered phase, and we set all initial link variables to unity. 
On the other hand, the circle (red) symbol denotes the ones started with “hot start”. The corresponding configuration is in the disordered phase, and the initial link variable is a random number. 
The square (black) symbol shows the result of the standard three-flavor QCD with the periodic boundary condition for spacial directions and the anti-periodic condition for temporal direction 
with the same values of $\beta$ and $\kappa$ as $Z_{3}$-QCD simulations. 

Now, let us look into the results in details.

Firstly, for the $Z_3$-QCD model, we find hysteresis in the range of $1.55 < \beta < 1.90$
depending on the initial conditions (cold or hot).
On the other hand, we find that there are no hysteresis in the data for the standard three-flavor QCD.
We note that the hysteresis is a signal of the first order phase transition.

Secondly, in the low-temperature phase, the magnitude of Polyakov loop is exactly zero for the $Z_3$-QCD model.
It originates in the existence of exact $Z_{3}$ center symmetry.
From these results, we argue that the $Z_3$-QCD model undergoes first-order phase transition, 
where the $Z_{3}$ center symmetry is spontaneously broken, 
while the standard three-flavor QCD undergoes the crossover transition.

\begin{figure}
\centering
\includegraphics[width=0.8\textwidth,clip]{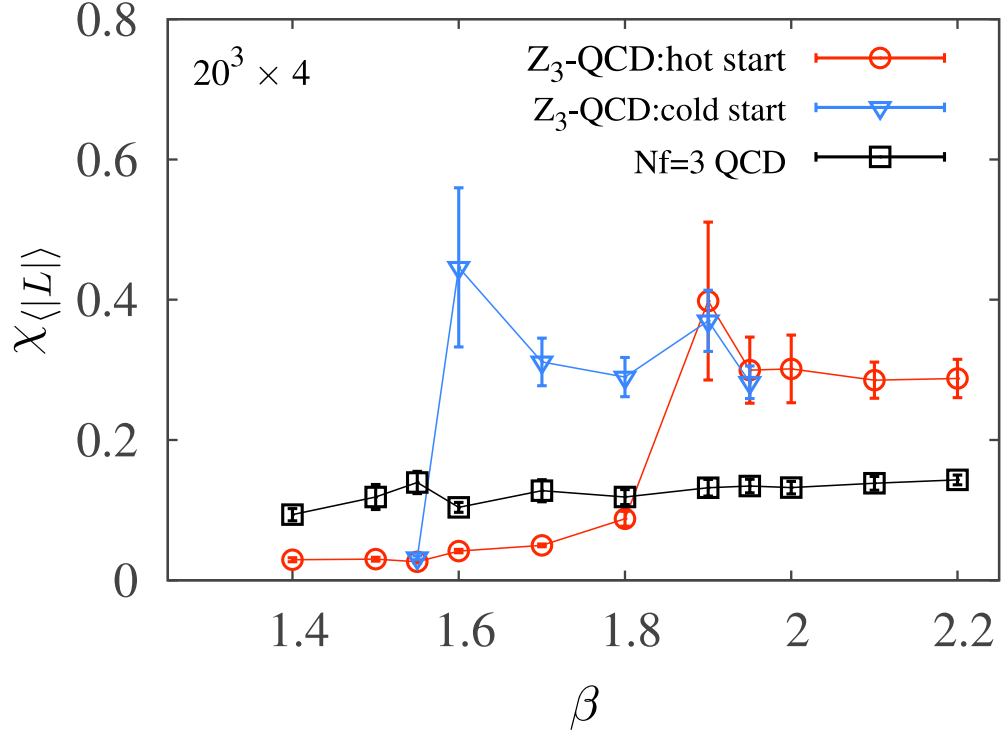} 
\caption{ 
Polyakov loop susceptibility $\chi_{\langle | L | \rangle}$ as a function of $\beta$ on $20^3 \times 4$ lattices.
Circle (red) and triangle (blue) symbols denote the data generated with hot and cold starts in $Z_3$-QCD model, respectively.
Square (black) symbol shows the data for standard three-flavor ($N_f=3$) QCD model.
}
\label{fig:chi-Ploop}
\end{figure}

Figure~\ref{fig:chi-Ploop} shows the Polyakov loop susceptibility defined by
\beq
\chi_{\langle | L | \rangle} \equiv V  \left[ \langle |L|^2 \rangle - \langle |L| \rangle^2 \right] . 
\eeq
The meaning of colors of symbols is the same as the one in Fig.~\ref{fig:Pb}.
The data for the $Z_3$-QCD model has a relatively clear signal of peaks around $\beta=1.60$ and $\beta=1.90$ for cold and hot starts, respectively.
The peak appears because of the co-existing states between the two phases associated with the phase transition~\cite{Fukugita:1990vu}.
On the other hand, the standard three-flavor QCD does not show a clear transition point. Actually, such a situation makes it difficult to determine the critical temperature in the physical QCD \cite{Borsanyi:2010bp, Bazavov:2011nk, Bhattacharya:2014ara}.

We also investigate the volume dependence of the critical $\beta$ and find small finite volume effect,
which we will not discuss further in the present paper.
The precise determination of the critical temperature in the large volume and continuum limits remains as a future work.

Our results on center symmetric phase transition indicates the following points:
\vspace{0.5cm}

(1) $Z_{3}$-QCD model on the lattice possesses exact $Z_{3}$ center symmetry.

(2) $Z_{3}$ center symmetry is spontaneously broken in the high-temperature phase.

(3) The order of the center phase transition is first.

\subsection{Chiral condensates}
Here, we investigate the chiral property in the finite-temperature $Z_3$-QCD mode, 
which is characterized by the chiral condensate.
In the chiral limit, the non-vanishing chiral condensate is 
a signal of the spontaneous chiral symmetry breaking.
As we discussed in Sec.~\ref{sec:chiral-Z3}, because of the twisted boundary condition, 
the symmetry breaking pattern is expected to be $U(1)^{2}_L \times U(1)^{2}_R \to U(1)^{2}_{\bar{V}}$ 
in the present model.
On the other hand, as we have discussed in Sec.~\ref{subsec:chiM}, the result of the chiral model implies 
that the flavor symmetry breaking due to the flavor-dependent twisted boundary condition gets manifest
only in the high-temperature phase~\cite{Kouno:2013mma}.
To investigate this conjecture, 
we will investigate the chiral condensate for each flavor independently.

Since our simulation is performed for heavy mass region $m_{\rm PS}/m_{V}=0.70$ 
and utilize the Wilson fermion,
no chiral phase transition takes place.
Nevertheless, we expect that, in the system with massive fermions, decrease of chiral condensate indicates effective and approximate restoration of chiral symmetry.
Actually, several lattice numerical simulations in these years 
reveal the chiral property for the realistic $N_f=2+1$ QCD at finite temperature
based on chiral condensate and its susceptibility as the effective order parameter.
We note that the recent lattice results using the staggered fermions and domain-wall fermions give 
the consistent chiral critical temperature~\cite{Borsanyi:2010bp, Bazavov:2011nk, Bhattacharya:2014ara} .

We consider the following flavor-diagonal expectation value of the subtracted chiral condensate~\cite{Giusti:1998wy,Umeda:2012nn,Hayakawa:2013maa} for each flavor~\cite{Bochicchio:1985xa,Itoh:1986gy,Aoki:1997fm} (we do not consider flavor-mixing condensate in the present paper.),
\beq
 \langle \bar{\psi}^f \psi^f\rangle_{\mathrm{subt.}} &=&  (2m_{\rm PCAC})(2\kappa)^2 \sum_{x} \langle P(\vec{x},t) P^\dag(\vec{0},0) \rangle.   \label{eq:def-chiral}  
\eeq
Here, $m_{PCAC}$ is partially conserved axial current (PCAC) mass, and $P$ denotes the pseudo scalar state defined by $P \equiv \bar{\psi}^f \gamma_5 \psi^f$, for the corresponding flavor ($f$) in LHS.
The PCAC mass is defined via axial Ward identity;
\beq
2m_{PCAC} =\frac{ \sum_{\vec{x}} \partial_4 \langle A_4 (\vec{x},t) P^\dag (\vec{0},0) \rangle }{ \sum_{\vec{x}} \langle P(\vec{x},t) P^\dag(\vec{0},0) \rangle },
\eeq
where $A_\mu$ corresponds to the axial vector current defined by $A_\mu =\bar{\psi}^f \gamma_5 \gamma_\mu \psi^f$. Here again, the label of flavor ($f$) is fixed.
The values of $m_\mathrm{PCAC}$ are summarized in Appendix \ref{sec:mass_summary}.

Before showing our results of numerical simulation for chiral condensates for each flavor, 
we comment on the notation of flavor in this study.
To fix the name of quarks, we firstly observe the complex phase ($\phi$) of Polyakov loop given as 
$L=|L|e^{i\phi}$ for each configuration.
As discussed in the previous section, in the center-symmetric phase the definition of the complex phase is meaningless since $\langle |L| \rangle =0$, while in the center broken phase we can define the value of $\phi$~\footnote{In principle the vacuum tunneling between three equivalent vacua makes difficult giving a clear definition of the phase. However, in our simulations, we did not encounter such difficulty.}.
Next, we measure the correlators in Eq.~(\ref{eq:def-chiral}) using the Dirac operator given in Eq.~(\ref{eq:def-Dirac}), where the link valuable has the boundary condition given by
\beq
U_\tau (\vec{x},\tau=N_\tau)=- e^{i \theta}U_\tau (\vec{x},0),
\eeq
where $\theta$ takes value $0$ or $\pm 2\pi/3$.
We define the flavor $u$ as in  
\beq
\phi + \theta = 0 \pmod{2\pi} \mbox{ for $u$-flavor}, 
\eeq
while the flavor $d$ and $s$ are defined with the following total phases; 
\beq
\phi + \theta &=& 2\pi/3 \pmod{2\pi} \mbox{ for $d$-flavor},\\ 
\phi + \theta &=& 4\pi/3 \pmod{2\pi} \mbox{ for $s$-flavor}.
\eeq

\begin{figure}
\centering
\includegraphics[height=8cm]{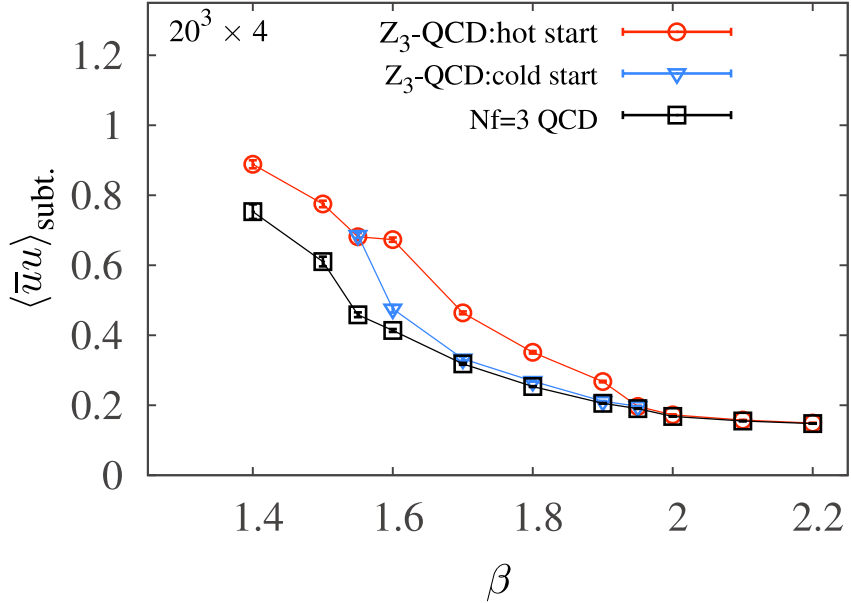} 
\caption{ 
$\beta$ dependence of the expectation values of subtracted chiral condensates $\langle \bar{u} u \rangle_{\mbox{subt.}}$ for $Z_{3}$-QCD and three-flavor QCD on $20^3 \times 4$ lattices.
Circle (red), triangle (blue) and square (black) symbols stand for those data associated with the hot start and cold start in $Z_3$-QCD model and the standard three-flavor QCD, respectively.
The error bar denotes the statistical error coming from the pseudo-scalar correlator in Eq.(\ref{eq:def-chiral}).
}
\label{fig:chiral-cond}
\end{figure}

Figure~\ref{fig:chiral-cond} shows the $\beta$ dependence of the chiral condensate for $u$-flavor of the $Z_{3}$- and three-flavor QCD.
Circle (red), triangle (blue) and square (black) symbols stand for those data associated with the hot start and cold start in $Z_3$-QCD model and the standard three-flavor QCD at finite temperature, respectively.
All results have common qualitative properties, 
where the chiral condensate gradually decreases as $\beta$ increases.
It is also notable that $\beta$ dependence of chiral condensates in $Z_3$-QCD model 
has hysteresis between the cold and hot starts as with that of the Polyakov loop.  
These results imply that effective restoration of the chiral symmetry is progressing from $\beta = 1.55$ to $\beta=1.95$ for the $Z_3$-QCD model.
These values of $\beta$ where the hysteresis exists are the same as those of the center phase transition.
Furthermore, the decreasing rate of chiral condensate in $Z_3$-QCD model is 
relatively larger than the one in the standard three-flavor QCD.

According to the arguments in Refs.~\cite{Roberge:1986mm, D'Elia:2002gd},
the partition function for QCD has $2\pi/3$ periodicity in the imaginary chemical potential, 
and it is expected that the chiral phase transition temperature in our model
is the same as that of the standard three-flavor QCD.
On the other hand, since the flavor-chiral symmetry in $Z_{3}$-QCD model in the chiral limit 
is broken to $U(1)^{2}_{L} \times U(1)^{2}_{R}$ due to the twisted boundary condition,
we may have the smaller number of Nambu-Goldstone modes than the usual three-flavor QCD,
which lifts the phase transition temperature in general \cite{Nagata:pc}.
(As we have discussed, the chiral effective model indicates that the full flavor symmetry is 
effectively preserved in the low-temperature phase, and in such a case
we should have the common number of Nambu-Goldstone modes.)
Our results in Fig.~\ref{fig:chiral-cond} indicate qualitatively the same chiral crossover temperature
in $Z_{3}$-QCD and three-flavor QCD, while, to be quantitative, 
the temperature in $Z_{3}$-QCD seems slightly higher than that in the three-flavor QCD. 
We do not yet have sufficient ingredients to conclude on this question.
Higher statistics and investigation of its susceptibility are 
necessary to determine the critical temperature in the massless limit.
It is of well-known difficulty to determine the critical temperature 
of the chiral phase transition~\cite{Meyer}.

We also note that the absolute values of the chiral condensates in $Z_{3}$- and three-flavor QCD are different in the low-temperature phase,
which may indicate the qualitative difference of the chiral property between the two theories.

\begin{figure}
\centering
\includegraphics[height=8cm]{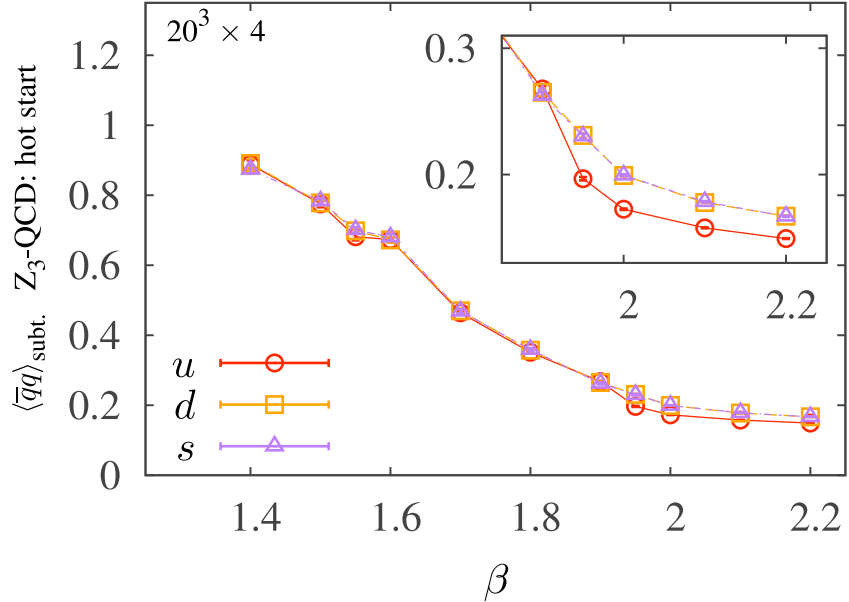} 
\caption{$\beta$ and flavor dependences of the expectation values of subtracted chiral condensates $\langle\bar{q}q \rangle$ for each flavor
in $Z_{3}$-QCD model. The lattice size is $20^3 \times 4$.
Circle (red), square (orange) and triangle (violet) symbols denote $u$-, $d$- and $s$-flavor generated with hot start, respectively.
}
\label{fig:chiral-flavor}
\end{figure}

Next, we focus on the flavor symmetry breaking in the high-temperature phase.
Figure~\ref{fig:chiral-flavor} shows the expectation values of chiral condensates for each flavor.
Here, circle (red), square (orange) and triangle (violet) symbols denote $u$-, $d$- and $s$-flavor generated with hot start, respectively.
Three components of chiral condensate are degenerate in the low-temperature phase.
On the other hand, in the high-temperature phase, there appears clear flavor symmetry breaking.
Two of them, whose total complex phase ($\phi + \theta$) are nontrivial, are degenerated 
because of the momentum shift of the twisted boundary condition.
It indicates that at least the $Z_3$ center of $SU(3)$ flavor symmetry, which commutes with the Cartan subgroup, is effectively preserved in the low-temperature phase,
while the breaking of this symmetry gets manifest in the high-temperature phase.
Although our simulation cannot fully verify the conjecture of the chiral model shown in (ii) 
of Sec.~\ref{subsec:chiM}, 
which states that the $SU(3)$ flavor symmetry is not affected by the twisted boundary condition 
in the low-temperature phase, the above result on the $Z_{3}$ flavor symmetry is consistent with this conjecture.
On the other hand, Fig.~\ref{fig:chiral-cond} shows the different values of chiral condensates 
between the $Z_{3}$ and three-flavor QCD, 
thus the two theories seem to have difference in chiral and flavor properties.
Further study is required to understand the structure of flavor and chiral symmetry in the $Z_{3}$-QCD model.

\section{Summary and Discussion}
\label{sec:summary}
In this work we numerically simulate the finite-temperature $Z_{3}$-QCD model on the lattice by introducing flavor-dependent twisted boundary conditions, with emphasis on center phase transition and its influence on the chiral properties.
We calculate the finite-temperature expectation value of Polyakov loop 
and the flavor-diagonal chiral condensates on $16^3 \times 4$ and $20^3\times 4$ lattices on the line of constant physics realizing $m_{PS}/m_{V}=0.70$.
We find out the first-order center phase transition at a certain temperature 
with the hysteresis depending on the initial conditions for configuration generation.
The chiral crossover transition takes place around the critical temperature of center transition, 
where it also has hysteresis in the same range as that of the center phase transition.
By comparing this approximate chiral restoration to that of the standard three-flavor QCD,
we find that the decrease of chiral condensate in $Z_{3}$-QCD is more rapid 
while temperatures of the two crossover transitions are almost the same.
We also obtain the result which supports the manifestation of flavor symmetry breaking 
due to the boundary condition in the high-temperature phase,
which was predicted in the study of the effective chiral model.

One of motivations for this work is study how the chiral condensate reacts to the center phase transition.
Our preliminary result indicates not only that the chiral condensate has rapid decrease
around the center critical temperature, but also that it has the hysteresis 
in the same range as that of the center phase transition.
Based on this result, we argue that the chiral and center properties at finite temperature 
have a strong correlation at least in the present model.
One possible reason for the correlation is that the first-order center phase transition might
work as a trigger to the rapid decrease of chiral condensate.
On the other hand, the argument in Ref.~\cite{D'Elia:2002gd} implies that the chiral critical temperature
in the present model is the same as that of the standard three-flavor QCD,
thanks to the $2\pi/3$ periodicity of imaginary chemical potential.
Our result on comparison of the two chiral (crossover) transition temperatures is consistent with this conjecture.
The precise study on the critical temperature in the $Z_3$-QCD model may help determine
chiral crossover temperature $T_c$ in the standard three-flavor QCD.

By comparing our results to those of the effective chiral model, we have supported the several conjectures 
based on the chiral model: {\it first-order center phase transition}, {\it latency of flavor symmetry breaking in the low-temperature phase}, and {\it manifestation of flavor symmetry breaking in the high-temperature phase}.
In particular, we have observed that $Z_3$ center part of the $SU(3)$ flavor symmetry seems to be intact
in the low-temperature phase, while its breaking becomes manifest in the high temperature phase, 
accompanying the spontaneous breaking of $Z_{3}$ center of the $SU(3)$ color symmetry. 
This is consistent with the conjecture of the effective chiral model, but is not sufficient to
draw a conclusion.
On the other hand, we could not find the specific jump of the chiral condensate right at the center critical temperature, which was seen in the chiral model as discussed in (iii) of Sec.~\ref{subsec:chiM}. 
Although our simulation is not performed in the chiral limit,
we may be able to interpret the jump seen in the chiral model as a model artifact and 
a remnant of the rapid decreasing of the chiral condensate seen in our simulation
since the cut-off effect may be visible at the high-temperature in the effective model.
Further study is required to clarify the flavor and chiral properties in the $Z_3$-QCD model.

\vspace{1cm}
 
For future works, we address following points.

{\it Towards a chiral limit}:
By approaching a chiral limit, the $Z_3$-QCD model has both exact center and exact chiral symmetries in presence of fundamental quarks.
We may be able to determine the both center and chiral critical temperatures using the exact order parameters, Polyakov loop and chiral condensate.
Therein, we can compare the two critical temperatures
and discuss relation of confining and chiral properties in more details.

{\it Towards smaller lattice and larger volumes}:
To obtain the critical temperatures, we have to take a continuum limit to remove a lattice artifact and take large volume limit to estimate finite volume effects.
Several works for the three-flavor real QCD near the physical points have been progressing, so that in principle it is doable at present.
Comparison with the critical temperatures between the $Z_3$-QCD and the standard three-flavor QCD in these limits must be interesting for understanding the center and chiral structure. 

{\it Towards topological objects (fractional instantons)}:
In the presence of center-symmetric Polyakov loop holonomy, or equivalently
the center-symmetric twisted boundary conditions, fundamental topological objects
become fractional instantons with fractional topological charge as $1/N$ with 
$N$ being the rank of gauge group \cite{Eto:2004rz, Bruckmann:2007zh}.
In our simulations, we expect that $1/3$ fractional instantons are present, and have 
influence on physical observables such as topological susceptibility.
It could be valuable to investigate the topological properties in relation to the recent interest
in the novel topological objects \cite{Unsal:2007vu, Argyres:2012vv, Dunne:2012ae}.

{\it Towards further application of gauge theory with twisted boundary conditions}:
Our result is of significance in a sense that we first observed the center first-order phase transition 
in the lattice QCD simulation with dynamical fundamental quarks by introducing the twisted boundary condition  (See also Appendix~\ref{sec:other-model}). It is notable that the only difference between the standard three-flavor QCD and our $Z_{3}$-QCD model is the {\it twisted boundary condition}, and this difference disappears in the zero-temperature limit. It means that, by choosing appropriate boundary conditions for quarks, we can realize an ideal situation that the confining/deconfining properties are well described by center symmetry and Polyakov loop. We consider that such a technique using the twisted boundary conditions helps understand broader topics on nonperturbative aspects of strong-coupling physics.

\acknowledgments
We are grateful to K.~Kashiwa, H.~Kouno and M.~Yahiro for their valuable comments 
and correspondences on their related work during the entire course of our study.
We would like to thank  P.~de~Forcrand, A.~Nakamura, H.~Suganuma and M.~\"{U}nsal for useful discussions and comments at the seminar in YITP, the workshop in Kyushu University and Lattice2015.  
We appreciate K.~Nagata's useful comments on the first version of the paper.
The authors thank the Yukawa Institute for Theoretical Physics,
Kyoto University.
Discussions during the YITP workshop YITP-T-14-03 on ``Hadrons and Hadron Interactions in QCD'' were useful to complete this work.
Numerical simulation for this study was carried out on Hitachi SR16000 
and IBM System Blue Gene Solution at KEK under its Large-Scale Simulation Program (No.~14/15-12) 
and Hitachi SR16000 at YITP.
E.~I. and T.~I. are supported in part by Strategic Programs for Innovative Research
(SPIRE) Field 5. 
T.\ M.\  is in part supported by the Japan Society for the 
Promotion of Science (JSPS) Grants Number 26800147.

\appendix

\section{The flavor-dependent b.c. as imaginary chemical potential}
\label{sec:Imch}
Consider the transformation where the fermion fields $\Psi_f$ in Eq.~(\ref{Z3BC}) are transformed as
\begin{equation}
\Psi_f \,\,\,\to\,\,\, e^{i(f-1)\tau/\beta }\Psi_f\,.
\label{BCfree}
\end{equation}
Then, the action (\ref{Action}) is translated into
\begin{equation}
S\,=\, \int d^{3}x \int_{0}^{\beta} d\tau 
\left[
\sum_{f}\bar{\Psi}_{f}(\gamma_{\mu}D_{\mu}+i(f-1)\gamma_{4}/\beta+m)\Psi_{f} \,+\, {1\over{2g^{2}}} \tr F^{2}_{\mu\nu} \right]\,.
\end{equation}
This is the theory with the flavor-dependent imaginary chemical potentials and the usual anti-periodic
boundary conditions Eq.~(\ref{ABC}).

\section{Comment on the other possible $Z_3$ symmetric theory}
\label{sec:other-model}

In the canonical ensemble,
it is known that the Polyakov loop 
has exact zero expectation value \cite{Kratochvila:2006jx},
which is similar to that of $Z_3$-QCD.
Here, we briefly discuss the difference between
the canonical ensemble and our $Z_3$-QCD model.

The grand canonical partition function can be written by
\begin{equation}
  Z_{\mathrm{GC}}(T,\mu)=\int [DU] e^{-S_g [\beta, U]} \det M(U,\mu),
\end{equation}
where $\mu$ denotes a quark (real) chemical potential.
Using $Z_\mathrm{GC}(T,\mu)$, 
the canonical partition function can be expressed as
\begin{equation}
Z_{C}(T,Q)=\int_{-\infty}^{\infty} d \left( \frac{\mu_I}{T}  \right) e^{-Q\mu_I/T} Z_{GC}(T,\mu=i\mu_I).
\end{equation}
with the quark number $Q$.

By using ($2\pi T/3$)-periodicity of the grand canonical partition function
as $Z_\mathrm{GC}(T,i(\mu_I + 2\pi T/3)) = Z_\mathrm{GC}(T,i\mu_I)$,
the canonical partition function can be expressed as 
\begin{equation}
  Z_{\mathrm{C}}(T,B)
  = \frac{1}{2\pi} \int_{-\pi}^\pi d
  \left( \frac{\mu_I}{T} \right)
  e^{-i3B \mu_I/T} Z_{\mathrm{GC}}(T,i\mu_I)
\end{equation}
for the baryon number $B (= Q/3)$, apart from a normalization factor.

Insertion of the imaginary chemical potential with 
  \begin{equation}
    \mu_I = \frac{2 \pi T k}{3} \quad \text{for} \quad k = 0, 1, 2,
    \label{}
  \end{equation}
can be translated by the center transformation ($z(k) \equiv e^{i2\pi k/3}$) acting only on the link variables in the fermion action,
  \begin{equation}
    U_4(\vec{x},x_4 = 0) \rightarrow z(k) U_4(\vec{x},x_4 = 0).
    \label{}
  \end{equation}
  Therefore, the Dirac determinant satisfies 
  \begin{equation}
    \det M(z(k)U, i\mu_I) = \det M(U,i\mu_I - i2\pi Tk/3).
    \label{}
  \end{equation}

  Using the periodicity of $Z_\mathrm{GC}(T,i\mu_I)$ and the above relation, 
  the canonical partition function is expressed as
  \begin{eqnarray}
    Z_\mathrm{C}(T,B)
    &=&  \frac{1}{2\pi} \int_{-\pi}^\pi
    d \left( \frac{\mu_I}{T} \right)
    e^{-i3B \mu_I/T} \nonumber \\
    && \times \frac{1}{3}\left[ Z_\mathrm{GC}(T,i\mu_I) 
    + Z_\mathrm{GC}(T,i\mu_I - i 2\pi T/3) 
    + Z_\mathrm{GC}(T,i\mu_I - i 4\pi T/3) \right], \nonumber \\
    &=&  \frac{1}{2\pi} \int_{-\pi}^\pi
    d \left( \frac{\mu_I}{T} \right)
    e^{-i3B \mu_I/T} \frac{1}{3}
    \int [DU] e^{-S_g[\beta,U]} 
    \sum_{k=0}^2 \det M(z(k)U, i\mu_I),
    \label{}
  \end{eqnarray}
  which means that $Z_\mathrm{C}(T,B)$ is the average of three center sector.
  It leads the exact zero expectation values of the Polyakov loop as
  \begin{equation}
    \left\langle L \right\rangle_{Z_{\mathrm{C}}(T,B)} 
    \propto 1 + e^{-i2\pi/3} + e^{-i4\pi/3} = 0.
  \end{equation}

On the other hand, the $Z_3$-QCD model introduces the flavor dependent
imaginary chemical potential.  
Thus, the partition function is
given by 
\begin{eqnarray} 
  Z_{Z_3\mathrm{\mathchar`-QCD}}(T) =\int [DU] e^{-S_g [U,\beta]}
 \prod_{k=0}^2 \det M(U,\mu=i 2 \pi k T /3).
\end{eqnarray} 
This partition function keeps center symmetry, 
since the integrand is invariant under the center transformation as follows:
\begin{eqnarray}
  && \det M(U,\mu=0) \det M(U,\mu=i2\pi T/3) \det M(U,\mu=-i 2\pi T/3), \nonumber \\
  &&\quad \rightarrow
  \det M(e^{i2\pi/3}U,\mu=0) \det M(e^{i2\pi/3}U,\mu=i2\pi T/3) 
  \det M(e^{i2\pi/3}U,\mu=-i 2\pi T/3), \nonumber \\
  &&\quad =
  \det M(U,\mu=-i 2\pi T/3) \det M(U,\mu=0) \det M(U,\mu= i 2\pi T/3). \nonumber \\
  \label{}
\end{eqnarray}

\section{PCAC mass and PS meson mass}
\label{sec:mass_summary}

\begin{figure}
  \centering
  \includegraphics[width=0.47\textwidth,clip]{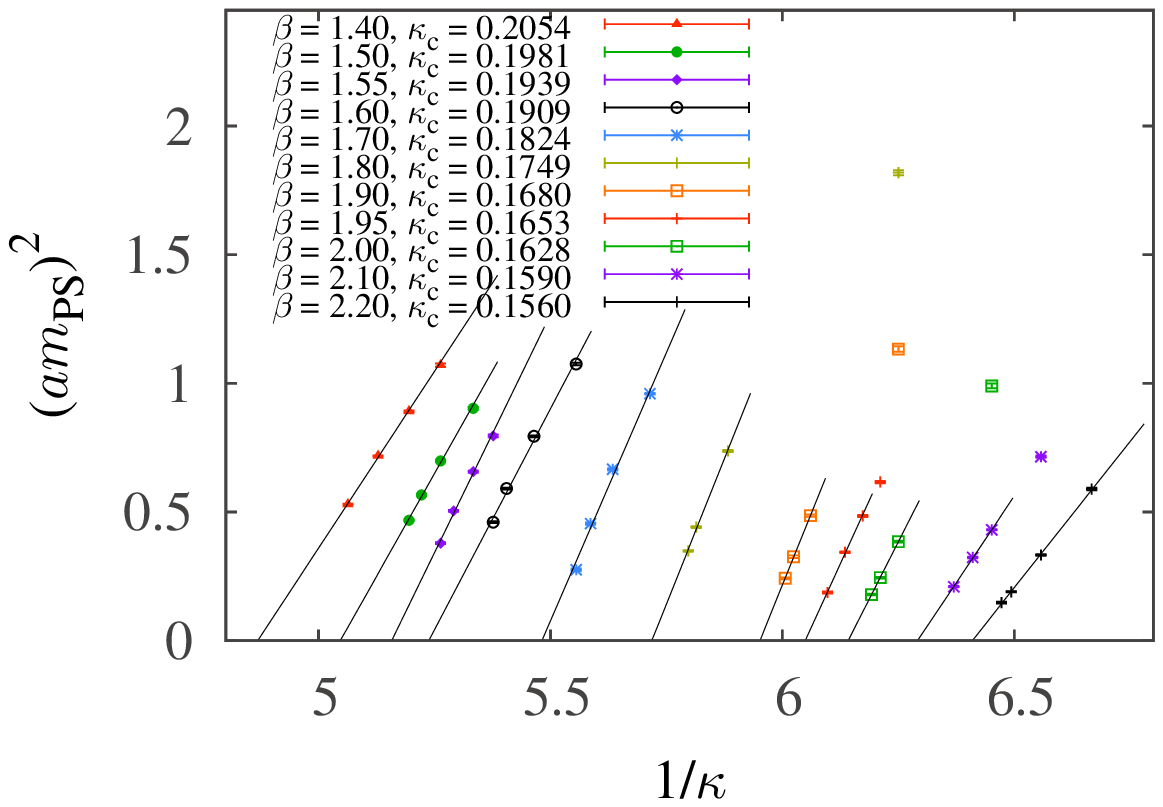}
  \includegraphics[width=0.47\textwidth,clip]{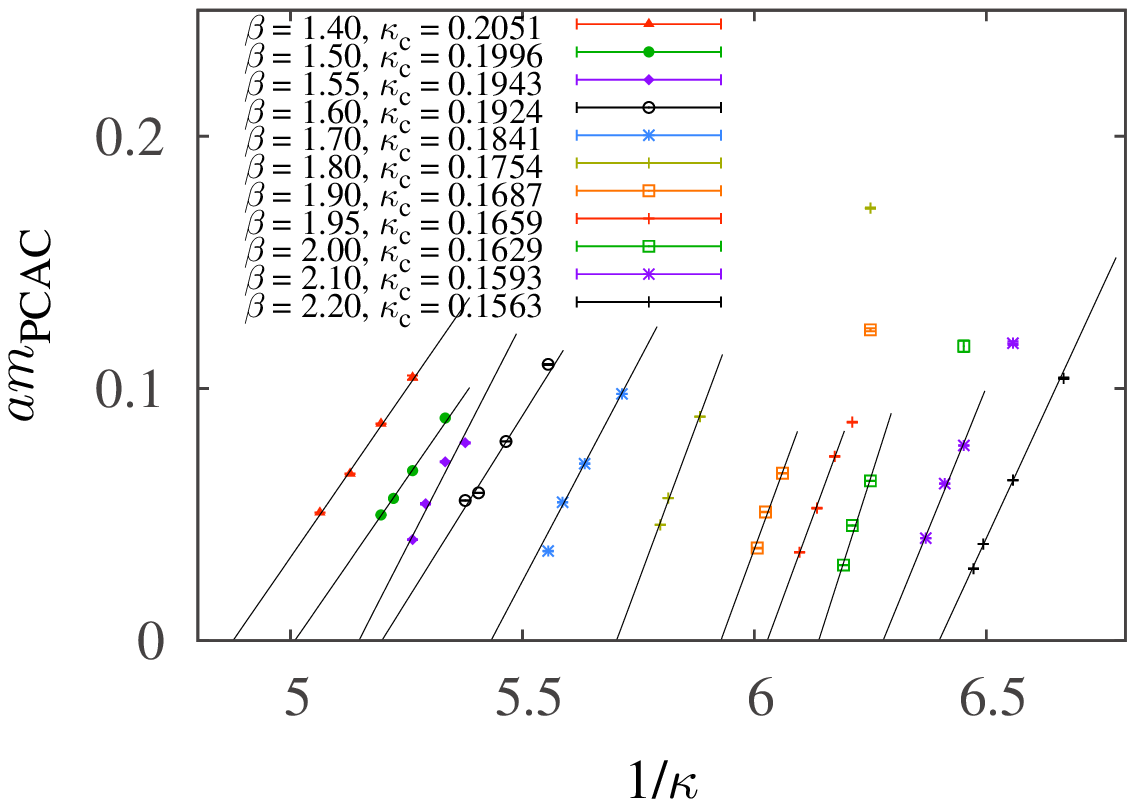}
  \caption{
    \label{fig:pcac_mpi}
    The hopping parameter $\kappa$ dependence PS and PCAC masses.
    Lines denote fit results using $(1/\kappa - 1/\kappa_c)$-type function.
  }
\end{figure}

Here, we summarize pseudo-scalar ($m_\mathrm{PS}$), 
vector ($m_\mathrm{V}$), their ratio
($m_\mathrm{PS}/m_\mathrm{V}$), and PCAC ($m_\mathrm{PCAC}$) masses
at the zero-temperature simulation using $16^4$ lattice 
in Table~\ref{tab:mass_summary_1} and \ref{tab:mass_summary_2}.
Figure~\ref{fig:pcac_mpi} shows the hopping parameter dependence 
of $m_\mathrm{PCAC}$ and $m_\mathrm{PS}^2$
with the fit results using $(1/\kappa - 1/\kappa_c)$.
For smaller masses, both are proportional to $(1/\kappa - 1/\kappa_c)$,
and they become zero at almost the same $\kappa$ up to $\mathcal{O}(am_\mathrm{PCAC})$,
which are expected behaviors of Wilson type fermion \cite{Aoki:1997fm}.
For the tuning of the mass parameter at $m_\mathrm{PS}/m_\mathrm{V} = 0.70$,
we interpolate $m_\mathrm{PS}$ and $m_\mathrm{V}$
as a function of $(1/\kappa - 1/\kappa_c)$ and determine 
the line of constant physics.

\begin{table}
  \centering
  \begin{tabular}{cccccc}
    \hline \hline 
    $\beta$ & $\kappa$ & $a m_\mathrm{PS}$ & $a m_\mathrm{V}$ & $m_\mathrm{PS}/m_{\mathrm{V}}$ & $a m_{\mathrm{PCAC}}$ \\
    \hline
    1.20 & 0.2000 & 0.9521(18) & 1.1883(37) & 0.8012(29) & 0.0901(1) \\
    1.20 & 0.2025 & 0.8704(11) & 1.1355(26) & 0.7665(20) & 0.0744(0) \\
    1.20 & 0.2050 & 0.7803(9)  & 1.0851(28) & 0.7191(20) & 0.0617(0) \\
    1.20 & 0.2100 & 0.5616(23) & 1.0003(108)& 0.5614(65) & 0.0360(3) \\
    \hline
    1.30 & 0.1900 & 1.1378(23) & 1.2988(42) & 0.8760(33) & 0.1257(2) \\
    1.30 & 0.1950 & 0.9960(11) & 1.2075(23) & 0.8248(18) & 0.0970(1) \\
    1.30 & 0.2000 & 0.8222(11) & 1.1027(29) & 0.7456(22) & 0.0724(0) \\
    1.30 & 0.2050 & 0.5873(13) & 0.9917(48) & 0.5922(32) & 0.0382(0) \\
    \hline
    1.40 & 0.1900 & 1.0345(31) & 1.2380(43) &  0.8356(38) & 0.1043(8) \\
    1.40 & 0.1925 & 0.9442(15) & 1.1690(25) &  0.8077(22) & 0.0860(1) \\
    1.40 & 0.1950 & 0.8472(11) & 1.1074(26) &  0.7650(21) & 0.0662(1) \\
    1.40 & 0.1975 & 0.7277(14) & 1.0429(41) &  0.6978(31) & 0.0508(1) \\
    \hline
    1.50 & 0.1875 & 0.9503(24) & 1.1668(44) &  0.8144(37) & 0.0883(2) \\
    1.50 & 0.1900 & 0.8358(17) & 1.0909(36) &  0.7662(30) & 0.0674(1) \\
    1.50 & 0.1915 & 0.7522(19) & 1.0399(35) &  0.7233(30) & 0.0564(1) \\
    1.50 & 0.1925 & 0.6838(16) & 1.0032(42) &  0.6816(33) & 0.0499(1) \\
    \hline
    1.55 & 0.1860 & 0.8919(31) & 1.1176(48) &  0.7980(44) & 0.0785(2) \\
    1.55 & 0.1875 & 0.8104(26) & 1.0627(58) &  0.7626(48) & 0.0709(2) \\
    1.55 & 0.1890 & 0.7098(24) & 1.0143(59) &  0.6998(47) & 0.0543(3) \\
    1.55 & 0.1900 & 0.6153(30) & 0.9308(56) &  0.6610(51) & 0.0401(2) \\
    \hline
    1.60 & 0.1800 & 1.0368(23) & 1.2159(37) &  0.8527(32) & 0.1096(2) \\
    1.60 & 0.1830 & 0.8910(21) & 1.1120(36) &  0.8013(32) & 0.0790(1) \\
    1.60 & 0.1850 & 0.7687(21) & 1.0326(34) &  0.7444(32) & 0.0586(1) \\
    1.60 & 0.1860 & 0.6784(23) & 0.9652(49) &  0.7029(43) & 0.0556(1) \\
    \hline
    1.70 & 0.1750 & 0.9794(18) & 1.1537(26) &  0.8489(25) & 0.0979(1) \\
    1.70 & 0.1775 & 0.8160(26) & 1.0329(48) &  0.7900(45) & 0.0702(1) \\
    1.70 & 0.1790 & 0.6746(33) & 0.9310(48) &  0.7246(51) & 0.0549(1) \\
    1.70 & 0.1800 & 0.5246(45) & 0.8351(60) &  0.6282(70) & 0.0355(2) \\
    \hline \hline
  \end{tabular}
  \caption{Summary of pseudo-scalar, vector, and PCAC masses for $\beta = 1.20 - 1.70$
  using $16^4$ lattice.}
  \label{tab:mass_summary_1}
\end{table}

\begin{table}
  \centering
  \begin{tabular}{cccccc}
    \hline \hline 
    $\beta$ & $\kappa$ & $m_\mathrm{PS}a$ & $m_\mathrm{V}a$ & $m_\mathrm{PS}/m_{\mathrm{V}}$ & $m_{\mathrm{PCAC}}a$ \\
    \hline
    1.80 & 0.1600 & 1.3487(36) & 1.4365(50) &  0.9389(41) & 0.1715(3) \\
    1.80 & 0.1700 & 0.8583(25) & 1.0399(28) &  0.8254(33) & 0.0888(1) \\
    1.80 & 0.1720 & 0.6644(26) & 0.8944(37) &  0.7428(42) & 0.0565(1) \\
    1.80 & 0.1725 & 0.5906(26) & 0.8392(40) &  0.7038(46) & 0.0459(1) \\
    \hline
    1.90 & 0.1600 & 1.0644(49) & 1.1807(60) &  0.9015(62) & 0.1233(6) \\
    1.90 & 0.1650 & 0.6968(34) & 0.8849(51) &  0.7874(59) & 0.0663(1) \\
    1.90 & 0.1660 & 0.5710(43) & 0.7905(43) &  0.7223(67) & 0.0510(1) \\
    1.90 & 0.1665 & 0.4920(31) & 0.7310(52) &  0.6731(64) & 0.0367(1) \\
    \hline
    1.95 & 0.1610 & 0.7852(28) & 0.9393(35) &  0.8359(43) & 0.0866(1) \\
    1.95 & 0.1620 & 0.6959(29) & 0.8642(39) &  0.8053(49) & 0.0731(1) \\
    1.95 & 0.1630 & 0.5861(31) & 0.7871(40) &  0.7446(55) & 0.0526(1) \\
    1.95 & 0.1640 & 0.4325(47) & 0.6754(51) &  0.6404(85) & 0.0351(1) \\
    \hline
    2.00 & 0.1550 & 0.9948(48) & 1.0942(91) &  0.9092(87) & 0.1167(23) \\
    2.00 & 0.1600 & 0.6203(28) & 0.7914(42) &  0.7838(55) & 0.0633(1) \\
    2.00 & 0.1610 & 0.4948(31) & 0.6996(35) &  0.7073(57) & 0.0457(1) \\
    2.00 & 0.1615 & 0.4231(37) & 0.6524(44) &  0.6485(72) & 0.0299(1) \\
    \hline
    2.10 & 0.1525 & 0.8455(25) & 0.9464(36) &  0.8934(43) & 0.1180(7) \\
    2.10 & 0.1550 & 0.6563(16) & 0.7934(27) &  0.8272(35) & 0.0774(1) \\
    2.10 & 0.1560 & 0.5683(24) & 0.7244(33) &  0.7845(49) & 0.0623(1) \\
    2.10 & 0.1570 & 0.4576(25) & 0.6423(34) &  0.7124(54) & 0.0406(1) \\
    \hline
    2.20 & 0.1500 & 0.7675(31) & 0.8677(31) &  0.8845(48) & 0.1040(3) \\
    2.20 & 0.1525 & 0.5769(27) & 0.7107(30) &  0.8117(51) & 0.0636(1) \\
    2.20 & 0.1540 & 0.4354(25) & 0.6283(41) &  0.6930(60) & 0.0383(0) \\
    2.20 & 0.1545 & 0.3845(41) & 0.6116(50) &  0.6287(84) & 0.0285(1) \\
    \hline \hline
  \end{tabular}
  \caption{Summary of pseudo-scalar, vector, and PCAC masses for $\beta = 1.80 - 2.20$
  using $16^4$ lattice.}
  \label{tab:mass_summary_2}
\end{table}

\end{document}